# Simultaneous 3D acoustic modulation and imaging of proton-induced nanodroplet vaporization for proton therapy monitoring

Sophie V. Heymans[1,2], Bram Carlier[3], Marcus Ingram[2], Kenneth Poels[4], Marc Desmaret[5], Nicolas Gerard[5], Alessandro Ramalli[6], Marc Fournelle[7], Tom Depuydt[4,8], Edmond Sterpin[8], Jan D'hooge[2] and Koen Van Den Abeele[1]

[1]Department of Physics, KU Leuven campus KULAK, 8500 Kortrijk, Belgium
[2]Department of Cardiovascular Sciences, KU Leuven, 3000 Leuven, Belgium
[3]Department of Imaging & Pathology, KU Leuven, 3000 Leuven, Belgium
[4]Department of Radiation Oncology, UZ Leuven, 3000 Leuven, Belgium
[5]Ion Beam Applications, 1348 Louvain-la-Neuve, Belgium
[6]Department of Information Engineering, University of Florence, 50139 Florence, Italy
[7]Fraunhofer-Institut für Biomedizinische Technik (IBMT), Sankt Ingbert, 66386 Saarland, Germany
[8]Department of Oncology, KU Leuven, 3000 Leuven, Belgium

Email: sophie.heymans@kuleuven.be

*This work was supported by the Research Foundation-Flanders (FWO) under the Junior Postdoctoral Fellowship grant number 12E0923N*

## Abstract

Superheated nanodroplets exposed to ionizing radiation vaporize into echogenic microbubbles, enabling ultrasound-based radiation therapy monitoring and proton range verification. However, at body temperature, decafluorobutane nanodroplets are insensitive to primary protons, and vaporize only when traversed by rare heavy secondary particles, limiting the *in vivo* applicability of this concept. We previously demonstrated that acoustic modulation at pressures well below the Acoustic Droplet Vaporization threshold can transiently increase nanodroplet sensitivity to radiation, enabling vaporization by primary protons. However, as acoustic modulation was applied by a focused single-element transducer, sensitization was restricted to a small region of the irradiated volume, and monitoring required an additional imaging probe. Here, we introduce a compact implementation that combines wide-field acoustic modulation and volumetric imaging on the same large-element matrix array, enabling simultaneous nanodroplet sensitization to protons and vaporization monitoring across the entire irradiated volume. The resulting 3D vaporization maps were fitted to the proton stopping distribution to estimate beam position and width parameters. At body temperature and starting from acoustic pressures as low as 122 kPa, we observed a transition from a response dominated by secondary particles to a response dominated by primary protons. The resulting vaporization peak at the proton range enabled sub-millimeter range estimation and detection of a 1.05 mm range shift with both accuracy and precision below 0.2 mm. These results demonstrate proton range verification and 3D treatment monitoring with acoustically-modulated decafluorobutane nanodroplets and represent an important step towards the envisioned clinical application.

## I. Introduction

Radiation therapy is, alongside chemotherapy and surgery, an essential weapon against cancer. Its success relies on the ability to accurately deliver dose to malignant tissue while minimizing exposure to surrounding healthy organs. In the past 30 years, proton therapy has steadily grown and established itself as an advanced radiation therapy modality [1]. In contrast to photons, which deposit dose along their entire path through the body, protons exhibit a finite, tunable range and concentrate most of their dose deposition near the end of their trajectory in the Bragg peak [2]. The steep distal dose fall-off and absence of exit dose offer substantial dosimetric advantages, but these can only be fully exploited if the proton range is accurately controlled. Small range errors can lead to substantial deviations between the planned and delivered dose distributions, affecting both tumor conformity and healthy tissue protection. Range uncertainties arise from both the treatment planning and delivery phases and currently necessitate the use of large safety margins and conservative treatment strategies, which prevent proton therapy from achieving its full potential [3].

Several techniques have been proposed for proton range verification and have reached different stages of development. The most mature approaches rely on secondary radiation generated by proton-induced nuclear interactions, either positron-emitting isotopes detected with PET imaging or prompt gamma rays detected with Prompt Gamma Imaging (PGI) [4]. Although both techniques have demonstrated clinical feasibility [5], [6], their widespread deployment remains hindered by system complexity, cost, uncertainties associated with the modelling of nuclear interactions in tissue, and the absence of simultaneous anatomical imaging for motion management [7], [8]. An alternative approach is ionoacoustics, also referred to as radiacoustic imaging, in which the rapid thermoelastic expansion induced by pulsed proton energy deposition at the Bragg peak generates acoustic waves detectable by ultrasound transducers [9], [10]. Unlike PET and PGI, ionoacoustics can be combined with conventional ultrasound, offering potential for real-time range verification in moving targets [11]. Recent studies, including a first proof-of-concept in patients [12], have demonstrated the feasibility of the technique for proton therapy [13], [14]. However, challenges related to signal sensitivity, limited field of view, and clinical integration continue to impede its clinical translation [10], [15], [16].

We recently introduced an ultrasound-based proton range verification approach leveraging superheated nanodroplets (NDs), also known as phase-change contrast agents. Upon exposure to ionizing radiation, NDs vaporize into echogenic microbubbles (MBs), whose spatiotemporal distribution correlates with the deposited radiation dose [17], [18]. While requiring a contrast injection, the proposed technique has the potential to achieve higher sensitivity than ionoacoustics, is compatible with both pulsed and continuous proton beams, and can be readily applied using conventional imaging arrays. The concept was recently validated in a small animal model [19].

The radiation-induced vaporization of metastable, superheated liquids is explained by the homogeneous nucleation and subsequent growth of vapor embryos along the tracks of charged particles, as described by Seitz's thermal spike theory [20]. In Seitz's model, vaporization occurs when a charged particle deposits at least the energy required to form a critical vapor embryo, $W_{\mathrm{tot}}$, within a distance comparable to the critical bubble radius $R_c$. This condition is expressed as a threshold on the track-averaged linear energy transfer (LET), $\langle\frac{dE}{dx}\rangle$, of charged particles, below which vapor embryos collapse and above which they grow into detectable vapor bubbles:

$$\langle\frac{dE}{dx}\rangle_{L_{\mathrm{eff}}} \geq \frac{W_{\mathrm{tot}}}{aR_c} \quad (1)$$

In this expression, the effective length $L_{\mathrm{eff}}$ over which energy must be deposited is related to the critical embryo size through the empirical nucleation parameter $a$, such that $L_{\mathrm{eff}} = aR_c$. Whereas $W_{\mathrm{tot}}$ and $R_c$ depend on the thermodynamic properties of the superheated liquid [17], [21], reported values of $a$ typically lie between 2 and 12 [21], [22], [23]. The LET threshold for vaporization is inversely related to the degree of superheat, *i.e.* the difference between the temperature of the liquid core and the boiling temperature corresponding to the ambient pressure [22], [24], [25]. For a given liquid, it increases with increasing temperature or decreasing pressure, thereby reducing the energy required to initiate vaporization. Of note, the thermal spike theory was developed for uncoated superheated liquids and does not account for the additional energy required to expand the viscoelastic shell of coated NDs.

Superheated NDs used in ultrasound imaging typically consist of a perfluorocarbon liquid core, owing to their biocompatibility and *in vivo* stability [26], [27], [28], [29]. Common formulations include perfluoropentane (PFP, $C_5F_{12}$, b.p. 28°C), decafluorobutane (DFB, $C_4F_{10}$, b.p. -2°C) and octafluoropropane (OFP, $C_3F_8$, b.p. -37°C). We previously showed that, at body temperature, the vaporization of DFB NDs exposed to proton beams is induced by rare high-LET secondary particles generated by nuclear interactions rather than by primary protons themselves [24]. Increasing the degree of superheat by raising the temperature to 50°C lowered the LET threshold sufficiently to enable vaporization by primary protons at the end of their range, resulting in a 25-fold increase in vaporization yield and a direct correspondence between the vaporization peak and proton range [18]. However, such temperatures are not compatible with *in vivo* applications, motivating the search for alternative strategies to increase the ND superheat.

Although replacing the DFB liquid core by OFP would yield a degree of superheat sufficient for proton sensitization at body temperature, OFP reaches its theoretical limit of superheat and becomes prone to spontaneous vaporization at approximately 38°C, which likely contributes to its reduced *in vivo* stability [30], [31]. As an alternative to modifying the ND liquid core, and inspired by findings in sono-photoacoustic imaging [32], [33], [34], we previously proposed the concept of acoustic modulation, in which the ND degree of superheat is dynamically controlled by an externally applied acoustic field. We hypothesized that NDs exposed to the wave rarefactional phase experience a transient decrease in ambient pressure, raising their degree of superheat and thereby lowering the LET threshold for vaporization. We validated this concept using a single-element focused transducer to modulate the pressure experienced by DFB NDs located at the distal end of the Bragg peak [35]. A 1.1 MHz acoustic modulation wave was applied at 72% duty cycle during continuous proton irradiation, while ND vaporization was monitored using a co-aligned linear array. Whereas neither proton irradiation nor acoustic modulation alone produced measurable vaporization, their combination enabled ND vaporization by primary protons, with a peak negative pressure (PNP) threshold of only 400 kPa at 37°C, well below that required for acoustic droplet vaporization (ADV) [35].

In this contribution, we address three limitations that currently hinder the clinical translation of acoustic modulation. First, rather than limiting acoustic modulation to the focal region of a single-element transducer, we extend acoustic modulation to a large three-dimensional field of view, enabling the sensitization of DFB NDs across the irradiated volume. This allows the spatial distribution of ND vaporization events to be used as a surrogate for the distribution of stopping protons and, consequently, for proton range verification. Second, we introduce a compact implementation in which a single matrix array is used both to generate the acoustic modulation field and to perform volumetric ultrasound imaging of ND vaporization. Requiring only a single acoustic window, this configuration minimizes the probe footprint and reduces interference with proton beam delivery. Third, instead of relying on prolonged acoustic sonication, we synchronize the delivery of short acoustic modulation pulses (~34 μs) with the proton pulses produced by a clinical synchrocyclotron, thereby reducing acoustic exposure and potential associated safety concerns. We validate the proposed 3D acoustic modulation strategy in a water phantom, determine the acoustic pressure threshold required for sensitization, and assess the proton range verification accuracy at body temperature.

## II. Materials and Methods

### A. ND synthesis

DFB NDs with a polyvinyl alcohol shell (PVA-DFB NDs) were prepared as described in [24] and stored at 4°C for maximally 2 days before the irradiation experiment. The ND size distribution was measured using Dynamic Light Scattering, leading to an intensity-weighted median diameter of 799 nm and polydispersity index of 0.3 [24]. The concentration of the DFB liquid core was quantified on the day of the irradiation experiment with Nuclear Magnetic Resonance spectroscopy (400 MHz Avance II, Bruker Biospin GmbH, Rheinstetten, Germany).

### B. Experimental setup

#### 1. *Clinical proton therapy system*

Proton irradiations were carried out in the treatment room of the proton therapy center of the Leuven university hospital (Particle, Belgium). The Particle center is equipped with a ProteusOne (Ion Beam Applications, Louvain-la-Neuve, Belgium) proton therapy system, including a superconducting synchrocyclotron (S2C2) delivering a proton pulse of ~10 µs at a pulse repetition frequency (PRF) of 1 kHz [36], [37].

#### 2. *Phantom description*

We irradiated phantoms made of PVA-DFB NDs diluted in deionized water at concentrations of 20 or 35 µM of DFB (Table I). The phantoms were enclosed in polymethyl methacrylate (PMMA) tanks of different dimensions depending on the incident beam energy (inner dimensions 85 × 70 × 80 $mm^3$ – small tank for 70 and 71 MeV irradiations or 129 × 114 × 104 $mm^3$ – large tank for 140 MeV irradiations). The PMMA wall through which the proton beam entered the water-based phantom had a thickness of 3 mm. The tank was placed on a heating plate equipped with temperature control to maintain the temperature at 37°C.

#### 3. *Ultrasound imaging*

An 11 × 11 matrix array (Fraunhofer IBMT, Sulzbach, Germany) [38] with a center frequency of 2 MHz (70% bandwidth) and an element pitch of 2.81 mm (square aperture of 30.9 $mm^2$) was positioned with the ultrasound azimuth direction (US-x) parallel to the proton beam direction (Fig. 1). An acoustically-transparent, 20 µm thick polyester window was placed between the matrix array and the phantom for acoustic coupling. The array was controlled by an ULA-OP 256 research scanner (X-Phase srl, Florence, Italy) [39].

#### 4. *Irradiation settings*

Phantoms were irradiated using monoenergetic, pristine proton beams at different energies (70, 71 and 140 MeV), in the presence or absence of acoustic modulation at different pressures. All experimental conditions are summarized in Table I. For the 140 MeV case, solid water slabs (SP34, IBA Dosimetry, Schwarzenbruck, Germany) were inserted in front of the phantom to ensure that the proton range was within the acoustic field of view of the matrix array. The proton beam current at the nozzle was varied to yield between $5 \times 10^5$ and $1 \times 10^6$ protons/pulse and the irradiation duration was controlled manually to be ~4 seconds, leading to clinically-relevant proton doses (peak doses at spot center and at the Bragg peak) between 3 and 9.4 Gy. Ultrasound acquisitions were controlled manually to capture the entire irradiation, starting ~1 second before and ending ~0.5 second after proton irradiation.

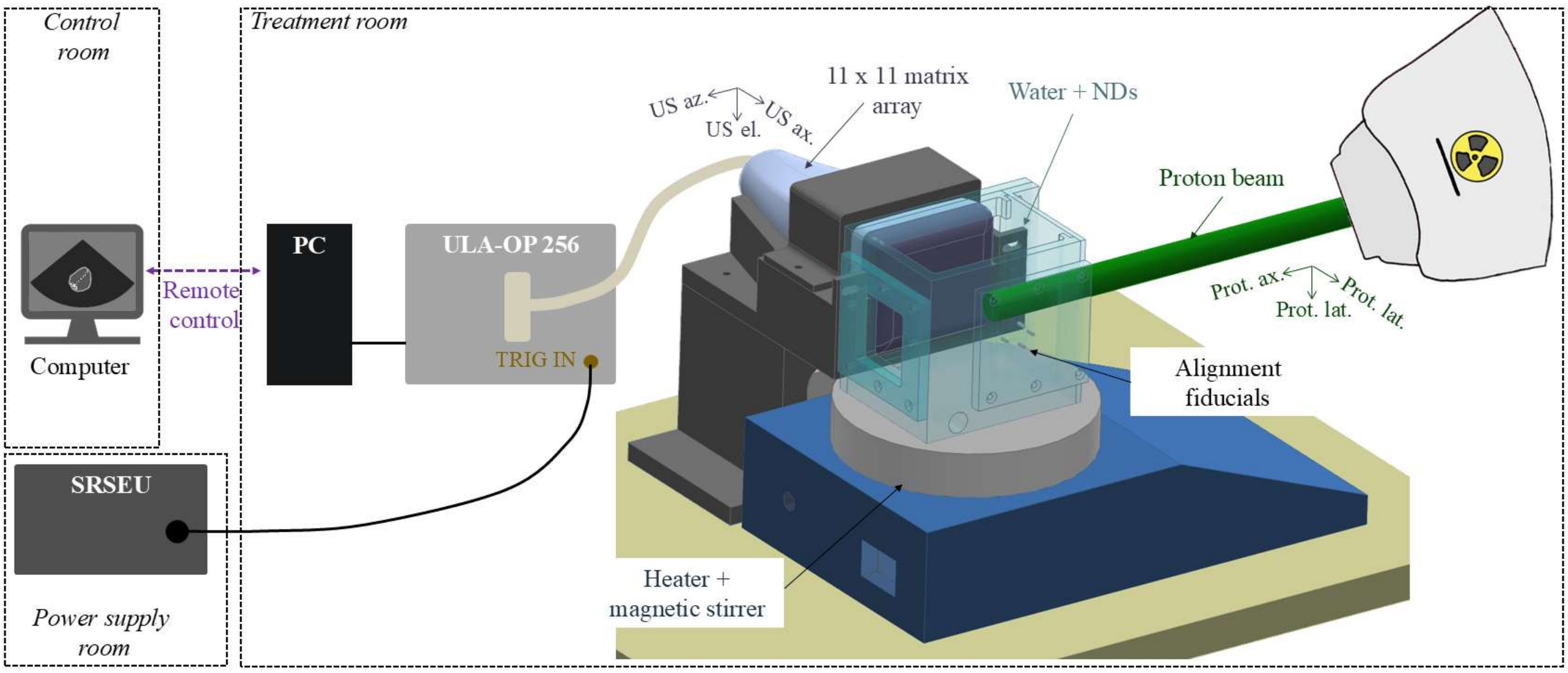

Fig. 1: Experimental setup used for synchronized proton irradiation and acoustic modulation of a solution of PVA-DFB NDs dispersed in water at body temperature. The matrix array used for acoustic modulation and interleaved imaging was controlled by an ULA-OP 256 research scanner, itself triggered by the SRSEU to ensure simultaneous arrival of the acoustic modulation wave with the proton pulse.

#### 5. *Synchronization between acoustic modulation and proton delivery*

To ensure that the arrival of the acoustic modulation pulse in the irradiated volume was synchronized with the arrival of each proton pulse, a signal sent by the Synchrocyclotron's Real-time and Sequencing Electronic Unit (SRSEU) at the same PRF as the proton PRF (1 kHz) was used as trigger input for the acoustic transmission of the ULA-OP 256. The fixed delay between the rising edge of this SRSEU signal and the proton pulse arrival was measured with a diode (Edge Detector, Sun Nuclear, Melbourne, FL, USA) on an oscilloscope (Virtual Bench, National Instruments, Austin, TX, USA) to be 279 µs. The transmission of the acoustic modulation pulse on the ULA-OP was then delayed such that the center of the pulse would reach the axial depth corresponding to the proton axis (45 mm) at the time of the proton pulse arrival (hence 279 µs after receiving the trigger input signal) (Fig. 2).

#### 6. *Interleaved volumetric imaging for vaporization monitoring:*

The acoustic modulation pulses transmitted at the same PRF as the proton PRF (1 kHz) were preceded by short imaging pulses for vaporization monitoring. The latency between reception of the SRSEU trigger signal and transmission of the acoustic modulation pulse was sufficient to enable the transmission (and full reception of echoes at depths up to the back wall) of one (large tank) or two (small tank) consecutive imaging pulses (Fig 2). Eight weakly diverging waves (virtual focus depth at 221 mm behind the probe, Fig. 2) positioned in an aperiodic pattern were compounded to yield volume rates of 250 Hz (small tank, two transmits per PRF) or 125 Hz (large tank, 1 transmit per PRF).

#### 7. *Registration between ultrasound and room coordinates*

The water tank was aligned with the proton beam using the in-room lasers and markings at the target beam position. To convert 3D ultrasound coordinates to 3D room coordinates, three translations and three rotation conversion factors were estimated by imaging four M2 set screw fiducials facing the matrix array at fixed, known positions. Briefly, ultrasound images of these fiducials were acquired with a 21 angle diverging waves sequence (with virtual foci forming an aperiodic spiral pattern [40]) and beamformed in 3D. Probe tilt angles in the elevation-axial (US-y/US-z) and azimuth-axial (US-x/US-z) directions were estimated by determining the tilt angle of the back wall, and the US-z position of the tank wall was used to translate axial ultrasound coordinates to room coordinates. After correcting for the US-y/US-z and US-x/US-z probe tilts, the azimuth and elevation positions of the tip of each fiducial were determined from three manually selected US-x/US-y ultrasound slices by evaluating their intensity-weighted center of mass coordinates. These coordinates were fitted to the known distances between the fiducials to determine the US-x/US-y probe tilt angle and the translational shifts between the US-x and US-y axes and corresponding room axes.

## C. Design and characterization of the 3D acoustic modulation field

#### 1. *Requirements for wide-field modulation*

To accurately recover the 3D proton distribution from the ND vaporization distribution, the acoustic modulation field should be as homogeneous as possible to ensure that all irradiated NDs have the same probability of proton-induced vaporization. Spatial pressure variations can affect vaporization in two ways: they determine whether a ND is sensitized at a given instant (negative pressure lower than the sensitization threshold), and they alter the duration for which the ND remains in a sensitized state. As a result, the observed vaporization distribution depends not only on the proton distribution but also on the local acoustic pressure field. In addition, the acoustic modulation field size should encompass the majority of the irradiated volume. In the ultrasound axial direction (US-z), this means that the modulation pulse length should be at least as long as the sum of the proton pulse length (~10 µs) and the proton beam width divided by the speed of sound.

Table I: Experimental conditions

| DFB concentration [µM] | Delivered protons | Beam energy [MeV] | Acoustic modulation pressure $P_{\mathrm{rms,av}}$ [kPa] | Frame rate [Hz] |
|---|---|---|---|---|
| | $4.38 \times 10^9$ | 70 | 265 | |
| 20 | $4.13 \times 10^9$ | | 0 | |
| | / | / | 265 | |
| | $2.70 \times 10^9$ | | 99 | |
| 20 | $2.78 \times 10^9$ | 70 | 185 | 250 |
| | $1.78 \times 10^9$ | | 265 | |
| | $4.07 \times 10^9$ | | 0 | |
| 20 | $4.30 \times 10^9$ | 71 | 265 | |
| | $3.99 \times 10^9$ | | 354 | |
| | $3.71 \times 10^9$ | 140 | 265 | |
| 35 | $5.43 \times 10^9$ | | 0 | |
| | / | / | 265 | |
| | $5.30 \times 10^9$ | | 99 | 125 |
| 35 | $5.15 \times 10^9$ | 140 | 122 | |
| | $4.48 \times 10^9$ | | 185 | |
| | $4.78 \times 10^9$ | | 265 | |

The use of a relatively long modulation pulse, combined with the large element size with respect to the wavelength, leads to strong grating lobes in the transmit pressure field of the 11 × 11 array. In addition, the array has limited focusing capabilities, which reduces the amount of steering that can be imposed to vary the pressure field. To compensate for the inhomogeneity of individual transmit fields, we chose to combine four different consecutive fields whose time-averaged (over four transmits i.e. 4 ms) acoustic distribution is homogeneous.

*2. Quantification of the pressure field homogeneity*

The spatial homogeneity of the acoustic modulation field was characterized using different metrics relevant in the context of acoustic modulation. These metrics were computed across a volume of 18 × 18 × 20 $mm^3$ (US-x × US-y × US-z) centered around the proton stopping distribution, and reported in %, normalized with respect to the maximum value: for a quantity $p$, the homogeneity metric $p_h$ represents:

$$p_h = \frac{\max(p) - \min(p)}{\max(p)} \times 100\ (\%) \tag{2}$$

We report on the homogeneity of the following metrics:

(1) The absolute lowest PNP across the four transmits ($P_{\text{min,all}}$), as the lowest pressure an ND might experience during proton irradiation:

$$P_{\text{min,all}}(x, y, z) = \min_{TX_{1\to 4}} PNP(|x| \le x_{\text{lim}}, |y| \le y_{\text{lim}}, z, TX) \tag{3}$$

with $x_{\text{lim}} = y_{\text{lim}} = 9$ mm.

(2) The root mean square (rms) negative pressure averaged across the four transmits ($P_{\text{rms,av}}$), which represents the pressure below which NDs are during 25% of the time (50% of the rarefactional phase).

$$P_{\text{rms,av}}(x, y, z) = \sum_{TX_i=1}^{4} \frac{PNP(|x| \le x_{\text{lim}}, |y| \le y_{\text{lim}}, z,\ TX_i)}{4} \tag{4}$$

(3) The percentage of the time spent by NDs below $P_{\text{rms,av}}$ ($t_{<P_{\text{rms,av}}}$) as a quantity representative of the vaporization probability. Indeed, if the value of $P_{\text{rms,av}}$ corresponds to the proton sensitization threshold, NDs are sensitized to protons 25% of the time (*i.e.* when the local negative pressure is below $P_{\text{rms,av}}$).

*3. Optimization of the spatial distribution of the acoustic modulation field*

We performed an exhaustive search using Field II simulations to find the combination of four transmit beams that yielded the most homogeneous acoustic modulation field. A focal depth of 250 mm was first selected empirically to provide sufficient PNP while maintaining a large field of view across the target depth range (35-55 mm). Pressure fields were then simulated for 121 transmit focus locations arranged on a square grid (steering angle < 8°). A 2D Tukey apodization window (13 points) was applied to reduce edge effects while maintaining 11 active elements in each direction. For each transmit, the pressure distribution was computed at five relevant depths with respect to the proton distribution (35, 40, 45, 50, and 55 mm). The optimal combination of four transmits was determined by minimizing a weighted measure of the spatial variation of $P_{\text{rms,av}}$ across the five depths:

$$\min_{\text{TX comb}} \left[ \sum_{z_i=1}^{5} w_i \times \left( \max\left(P_{\text{rms,av}}(x, y, z_i)\right) - \min(P_{\text{rms,av}}(x, y, z_i)\right) \right] \tag{5}$$

with higher weights assigned to the depth corresponding to the center of the proton beam axis (45 mm). The resulting acoustic modulation field was subsequently validated using a calibrated 0.5 mm needle hydrophone (Precision Acoustics, Dorchester, U.K.).

*4. Acoustic modulation pressure threshold determination*

To determine the acoustic modulation pressure threshold for proton sensitization at body temperature, the acoustic pressure was varied across a wide range using both different ULA-OP output voltages and pulse-width modulation. Indeed, the relatively long (~34 µs) modulation pulse was transmitted using the three-level pulsers of the ULA-OP, which limited the minimum transmitted voltage to 25 V. Pulse-width modulation was implemented as described in [41] to further reduce the output pressure, and the obtained acoustic pressures for different pulse-width modulation scenarios were calibrated using hydrophone measurements.

## D. Vaporization events detection and localization

The acquired radiofrequency channel data were IQ demodulated and beamformed offline using delay-and-sum weighted by the amplitude coherence factor [42]. To account for the limited angular sensitivity of the large transducer elements, dynamic receive apodization with an F-number of 1.2 was applied. The resulting 3D+time datasets were then processed using an in-house ultrasound localization microscopy (ULM) pipeline.

First, the beamformed volumes were high-pass filtered along the temporal dimension to enhance fast changes associated with newly formed MBs due to vaporization and suppress stationary or slowly-moving bubbles. Next, MBs were detected on each frame by identifying local intensity maxima above a fixed threshold. As the pulse-echo response of the array and the use of dynamic receive apodization introduced spatial variations in pixel intensity, the beamformed data were normalized using a Field-II calibration of the spatial response of a point target, before the MB detection step. Detected MBs were interpolated on a finer spatial grid and super-localized using variable-width Gaussian PSF fitting in azimuth and elevation, and using the leading edge corresponding of 20% of the peak value in the axial direction [43]. The MBs were then tracked over time using the Hungarian algorithm [44], and the initial position of each track was retained as the position of a vaporization event. This approach assumes that each track originates from a ND that vaporizes into a MB and then flows in the sample until it either leaves the 3D field of view or dissolves. To exclude pre-existing MBs entering the field of view, all detected vaporization events located within 2 mm of the volume boundaries were discarded. The obtained set of positions formed a 3D + time vaporization map.

The objective of the 3D ULM pipeline was to detect as many vaporizations as possible (as the precision for range verification increases with the amount of events) while minimizing the amount of false detections. The number of detected events was highly sensitive to (i) the fixed dB threshold for MB detection, (ii) the signal-to-noise ratio of the acquired data as well as the amount of background MBs, and (iii) how efficiently the temporal filter isolates vaporizations from noise and background MBs. The latter was phantom-dependent, as the use of acoustic modulation led to downward MB flow due to the acoustic radiation force; as a result, the temporal filter settings were phantom-dependent. As the ultrasound frames could be temporally aligned with the beam delivery using the irradiation log files, the number of (false or spontaneous) vaporizations taking place before or after irradiation could be quantified. To ensure sufficient false positives rejection, the fixed dB threshold was set such that at most 5% of all localized vaporizations took place outside of the irradiation time.

## E. Comparison between vaporization maps and the delivered proton distribution

ND vaporization maps were temporally aligned with the beam delivery using the irradiation log files, allowing to remove the few vaporizations (max 5% as explained above) taking place outside of the irradiation duration. Next, the vaporization maps ultrasound coordinates were converted to room coordinates as described in section II. B7., and compared to the 3D distribution of the proton stopping positions, which was computed combining beam measurements with analytical models as described below and in [45], [46].

### *1. Proton beam model*

The proton depth-dose (PDD) distribution was obtained by applying an analytical fit [47] to measurements performed with an ionization chamber in water (StingRay, IBA Dosimetry, Schwarzenbruck, Germany). The lateral distribution of the proton beam was modelled as a 2D Gaussian, using the measured entrance lateral spot size (Lynx PT, IBA Dosimetry, Schwarzenbruck,

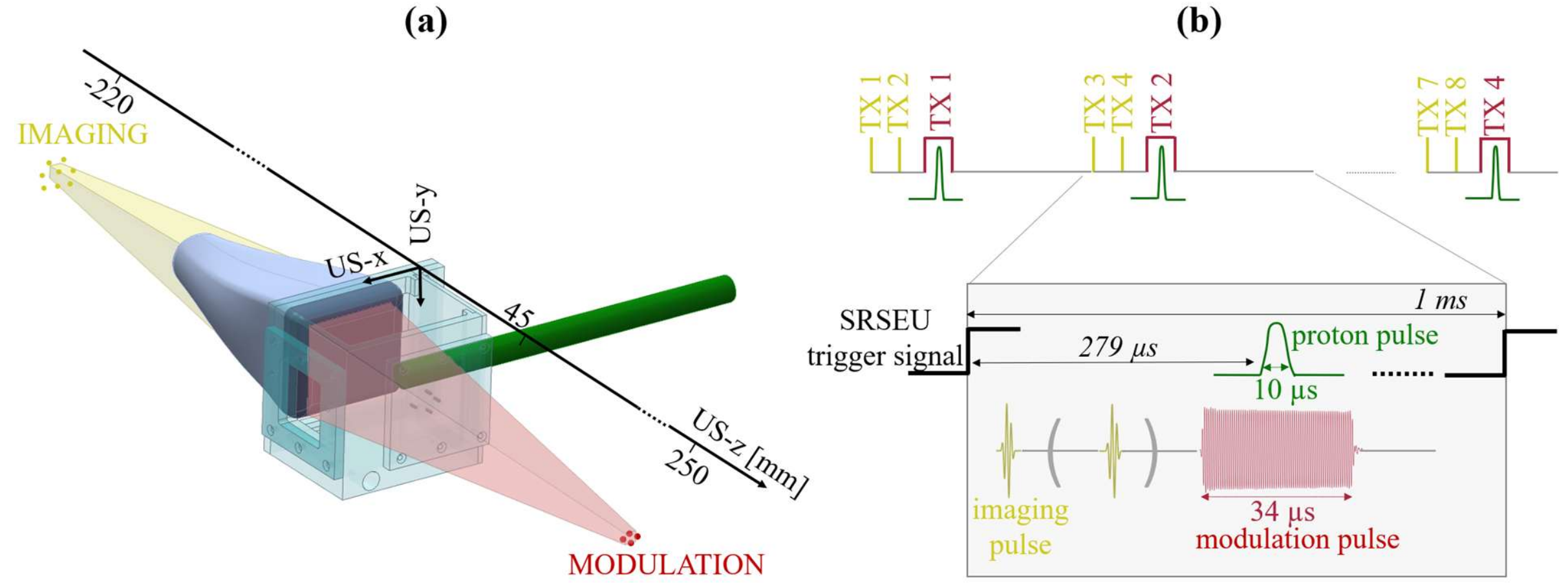

Fig. 2: (a) Schematic of the transmit strategy for 3D ultrafast imaging with slight diverging waves (yellow) and wide-field acoustic modulation (red). (b) Temporal sequencing of the imaging and modulation pulses: the ULA-OP transmits 2 (small tank) or 1 (large tank) imaging pulse (yellow) followed by a longer modulation pulse (red) timed to arrive at the center of the irradiated area (depth of 45 mm) at the same time as the proton pulse reaches the sample. The length of the modulation pulse is sufficient to ensure that the irradiated field is acoustically modulated for the entire duration of the proton spill (10 µs). The sequence is repeated at the proton PRF (1 kHz) using four different fields for acoustic modulation (TX1-4) and eight different virtual source positions for imaging (TX1-8).

Germany) and applying the Highland formula to model spot spreading in the water phantom [2], [48]. The proton range ($R_{80}$) in the phantom was adjusted accounting for the water-equivalent thickness (WET) of the solid water plates and PMMA phantom entrance wall [49]. To separate the contribution of primary protons and high-LET secondary particles to the total number of vaporizations, we computed the 3D distribution of primary protons stopping due to Coulombic interactions at the end of their range (vaporization by primary protons), $f_{\text{prim}}$, and the 3D distribution of protons lost due to nuclear interactions (vaporization by the produced secondary particles), $f_{\text{sec}}$. The former was modelled as a Gaussian peaking at the range position (being, by definition, the position at which half of the protons stopping due to Coulombic interactions have stopped [2]) and with a dispersion ($\sigma_x$) given by the PDD analytical fit [47], while the latter was derived from the total non-elastic interaction cross-section of protons in oxygen [50].

2. *Range fitting and estimation of the fraction of vaporizations due to primary protons*

A maximum log-likelihood estimation algorithm was used to fit the vaporization coordinates in the beam direction to a mixture model of the 1D projected distributions of protons stopping due to Coulombic interactions (primaries) and protons stopping due to nuclear interactions (secondaries):

$$p(x_i|w_{\text{prim}}, \mu) = (1 - w_{\text{prim}})f_{\text{sec}}(x_i - \mu_x) + w_{\text{prim}}f_{\text{prim}}(x_i - \mu_x) \tag{6}$$

where $x_i$ are the beam axial (US-x) coordinates of the vaporization events, $\mu_x$ is the translational shift between the observed vaporization distribution and the theoretical particle distributions, and $w_{\text{prim}}$ is the relative contribution of primary protons to the total number of vaporizations. The latter was left to vary as it depends on the ND degree of superheat, and hence the applied acoustic modulation pressure. The objective was to estimate the parameters $\boldsymbol{\theta} = \{w_{\text{prim}}, \mu_x\}$ given an observed set of $n$ vaporization coordinates $x_i$. The maximum a-posteriori estimation is performed by minimizing the negative log-posterior:

$$\widehat{\boldsymbol{\theta}} = \arg\min_{\boldsymbol{\theta}} \left( -\sum_{i=1}^{n} \ln p(x_i|\boldsymbol{\theta}) \right) \tag{7}$$

The optimization was performed with constrained nonlinear optimization (Matlab's *fmincon*) for multiple random initializations.

3. *Estimation of the lateral shifts*

The same approach was used to estimate the shifts between the proton distribution and the vaporization events coordinates transverse to the proton beam. In this case, only the distribution shift parameter $\mu_y$ or $\mu_z$, referring to the US-y and US-z directions, respectively, was retained, as contributions from primary and secondary particles were combined when projecting vaporizations in the beam lateral directions.

4. *Estimation of the beam width*

In addition to the translational shifts, the width of the vaporization profiles was also fitted to a Gaussian distribution in each direction to evaluate the difference between the vaporization distribution width and the true beam width.

To evaluate the uncertainties on the fitted parameters, the 95% confidence intervals for all the estimated parameters ($w_{\text{prim}}, \mu_x, \mu_y, \mu_z, \sigma_x, \sigma_y, \sigma_z$) were obtained through Monte Carlo simulations.

5. *Estimation of the vaporization yield by primary protons*

The number of protons delivered to each phantom was retrieved from the irradiation log files, and the beam model described in section II. E1. was used to obtain the number of primary protons stopping in the acoustic field of view due to nuclear interactions and due to Coulombic interactions. After fitting of equation (7), vaporization events attributed to high-LET secondary particles were subtracted from the total number of vaporizations, yielding the number of events attributed to vaporization by primary protons. This number was normalized by the ND concentration in the phantom and the proton fluence to provide the vaporization yield, and compared to theoretical expected values assuming that the NDs are sensitive to primary protons using a previously proposed analytical model [45].

## III. Results

### A. Characterization of the acoustic modulation pressure field

Field II simulations identified four optimal virtual focus positions on a square centered on the imaging axis (± 6.3 mm, 0) and (0, ± 6.3 mm), with a focal depth of 250 mm. The normalized PNP spatial distribution of each transmit field is shown in Fig. 3a-d across US-x and US-y at 45 mm depth (*i.e.* the center of the proton axis); hydrophone measurements are overlaid on simulations and delimited by the dashed black rectangles. Fig. 3e represents the global PNP ($P_{\text{min,all}}$) and shows that combining these four transmits beam improved the PNP homogeneity across a wide area. The maximum spatial variation of $P_{\text{min,all}}$ was 31% experimentally (hydrophone) and 33% in simulations, while the maximum variation of $P_{\text{rms,av}}$ was 22% experimental or 28% in simulations. The spatial distribution of $P_{\text{rms,av}}$ at the five considered depths is shown in Fig. 4a-e, again comparing hydrophone measurements (dashed rectangles) with Field II simulations. Fig. 5a-b compares the percentage of the time spent by NDs below the $P_{\mathrm{rms,av}}$ ($t_{<P_{\mathrm{rms,av}}}$) at 45 mm depth with the spatial distribution of the stopping protons in the US-x and US-y directions for two beam energies, while the 1D profiles shown in Fig. 4c and d illustrate the homogeneity of the pressure field in the US-y and US-z directions, respectively. 58.5% of the primary protons at 70 MeV, and 78% of the primary protons at 140 MeV stopped within the acoustic modulation volume defined above (18 × 18 × 20 $mm^3$). The mean percentage of time spent at negative pressures below the $P_{rms}$, $t_{<P_{\mathrm{rms,av}}}$, varied by up to 45% experimental and 58% in simulations. The temporal trace of the acoustic modulation pulse is visible in Fig. 2b; the pulse length of 34 µs corresponds to a physical length of 5.2 cm in water, which is sufficient to cover 95% of the Gaussian beam width ($\sigma_z$= 7.94 mm) for the duration of the proton pulse (~10 µs).

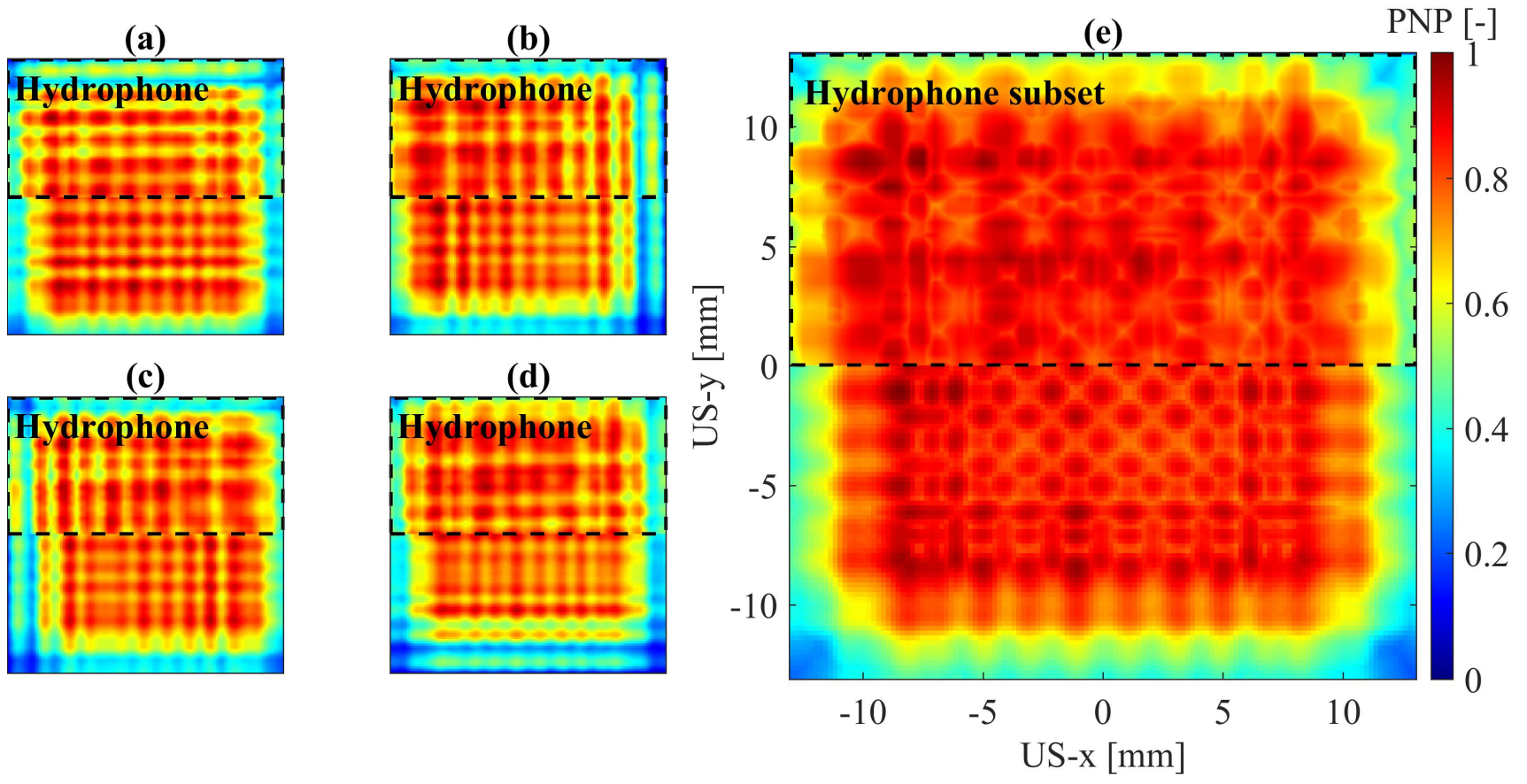

Fig 3: (a)-(d): PNP distribution (azimuth – elevation) for the four single transmits at a depth (US-z) of 45 mm. (e): distribution of $P_{min,all}$ across all transmits (azimuth-elevation) at a depth (US-z) of 45 mm. Dashed rectangles correspond to hydrophone measurements while the remainder is the output of Field II simulations.

## B. Acoustic modulation enables nanodroplet vaporization at body temperature across the irradiated volume

Fig. 6 shows the 3D distribution of localized vaporization events (dots) overlaid on the proton stopping distribution for 70 MeV (a) and 140 MeV (b) irradiations, comparing irradiated phantoms without acoustic modulation (grey), irradiated phantoms with acoustic modulation ($P_{\mathrm{rms,av}}$ = 265 kPa, green), and unirradiated control phantoms with acoustic modulation (yellow). At both proton energies, in the absence of acoustic modulation, the majority of the vaporizations were localized proximal to the proton range and followed the distribution of protons stopping due to nuclear interactions (gray color map). In contrast, when acoustic modulation was applied, most events were located at the proton range and followed the distribution of primary protons stopping due to Coulombic interactions (green contour lines). Non-irradiated phantoms did not exhibit significant droplet vaporization, which confirms that the acoustic modulation pressures were below the ADV threshold.

## C. The ratio between vaporizations due to secondary particles and primary protons can be modulated by the acoustic pressure

To evaluate the effect of the acoustic modulation pressure on the ND vaporization response to primary protons, the same phantom was irradiated several times using increasing modulation pressures (Table I). The fitted vaporization profiles in the direction of the proton beam (US-x) are displayed in Fig. 7 for 70 MeV (a) and 140 MeV (b) irradiations. For both energies, the vaporization response shifted from a uniform plateau proximal to the proton range to a plateau followed by a Gaussian profile coinciding with the stopping distribution of primary protons (green profile) peaking at the proton range (green dotted line). Fig. 8 shows the corresponding evolution of the fitted $w_{\mathrm{prim}}$, *i.e.* the fraction of vaporizations attributed to primary protons stopping at the end of their range, at 70 MeV (a) and 140 MeV (b). A value of $w_{\mathrm{prim}}$ equal to 0 indicates that all vaporizations are induced by secondaries, while a value of 1 indicates that vaporizations are exclusively induced by primary protons stopping at the end of their range. Different marker shapes correspond to different phantoms, and the displayed error bars correspond to the computed 95% confidence intervals on the estimated values. Both panels illustrate the gradual transition from a vaporization response dominated by secondary particles to a vaporization response dominated by primary protons as the modulation pressure increases.

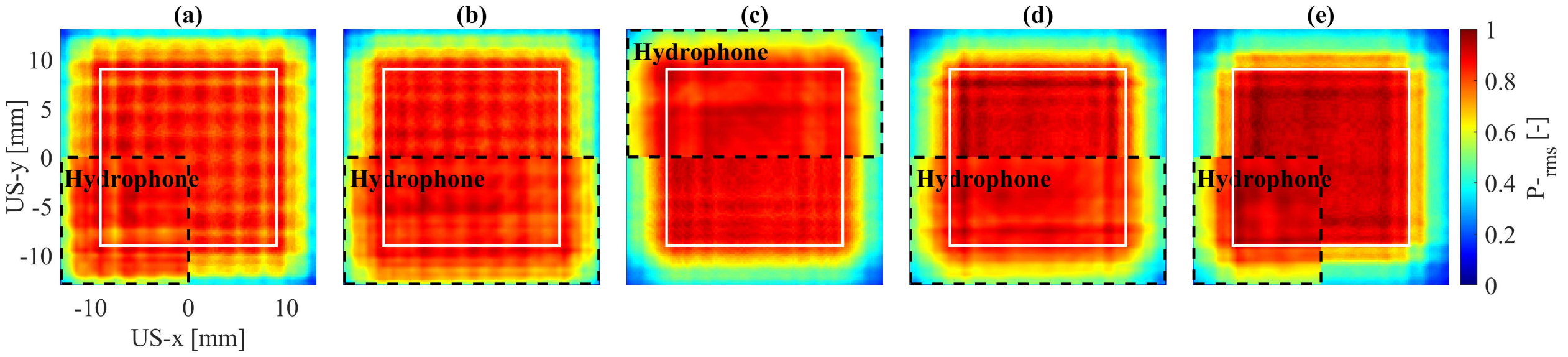

Fig 4: Spatial distribution of the rms pressure averaged across the 4 transmit events, $P_{rms,av}$, at 5 relevant axial depths with respect to the proton distribution: 35 mm (a), 40 mm (b), 45 mm (c), 50 mm (d) and 55 mm (e). Dashed rectangle inserts indicate hydrophone measurements while the rest of the maps are Field II simulations. The white rectangle represents the region considered when reporting the field homogeneity metrics.

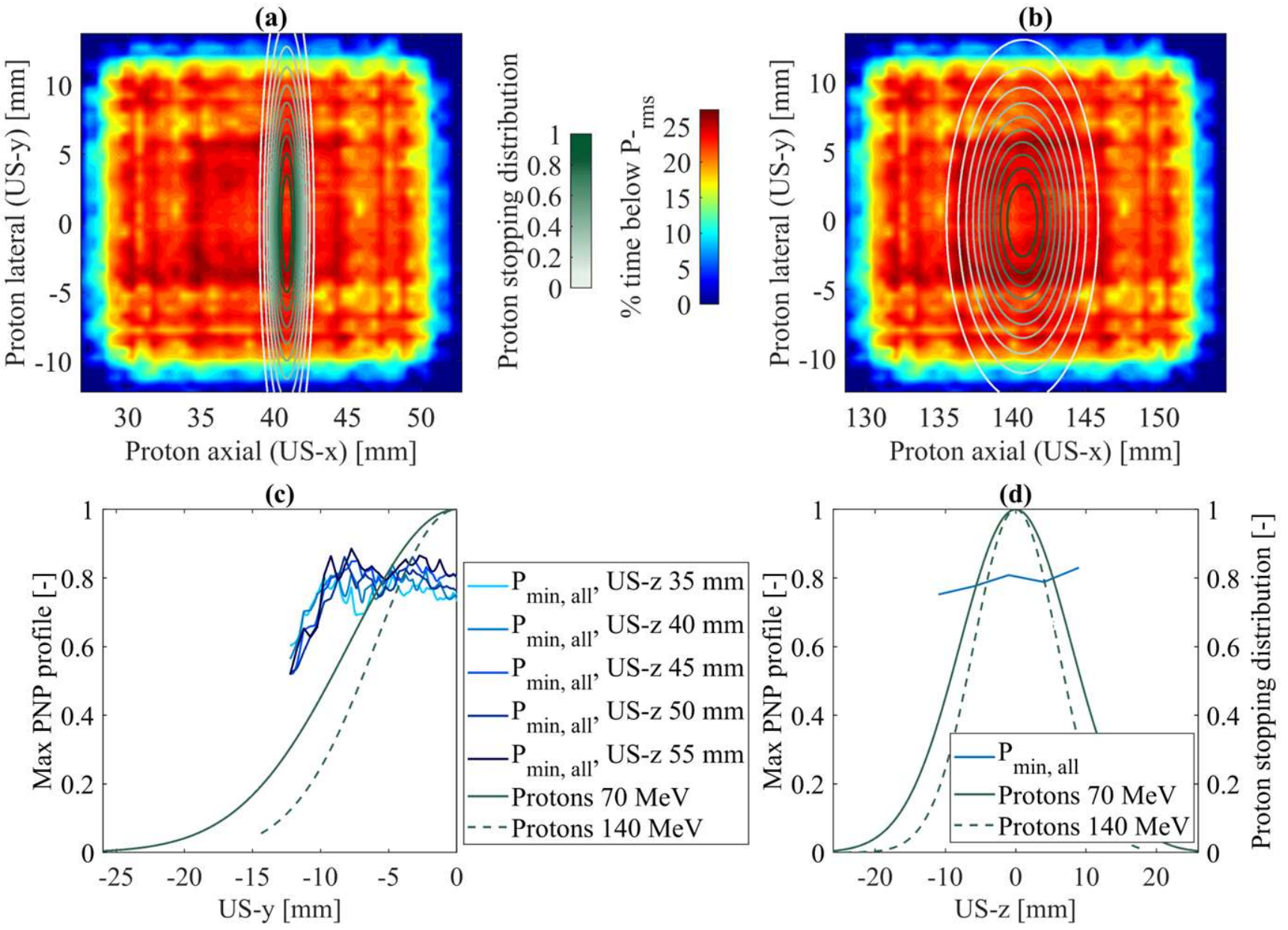

Fig 5: Spatial distribution of $t_{<P_{\mathrm{rms,av}}}$ at an axial depth of 45 mm (center of the beam axis), with the proton stopping distribution overlaid for (a) 70 MeV and (b) 140 MeV protons. 1D lateral profile of the proton stopping distribution at the two energies of interest compared to the acoustic modulation field size in the ultrasound elevation (US-y) direction (c) and axial (US-z) directions (d).

### D. Acoustic modulation enables the accurate detection of a 1 mm shift in proton range

To illustrate the impact of acoustic modulation on ND-based range verification performances, phantoms were irradiated at different energies of 70 and 71 MeV, causing a theoretical range shift of 1.05 mm. In the absence of acoustic modulation, the range shift estimated from the vaporization maps of two phantoms irradiated at these different energies was 2.82 mm. When acoustic modulation with $P_{\mathrm{rms,av}}$ = 265 kPa was applied during irradiation, we obtained a mean estimated range shift of 1.21 mm (*i.e.* within 0.16 mm of the theoretical range shift) with a standard deviation of 0.12 mm (computed using four combinations of phantoms , *i.e.* two phantoms at each energy).

### E. Volumetric imaging of the nanodroplet vaporization response for 3D treatment verification

The spatial distribution of vaporization events was fitted to the reference proton stopping distribution to estimate the translational shifts $\mu$ (distance-to-agreement) and Gaussian beam widths $\sigma$ in all three directions. The resulting parameters are shown as a function of acoustic modulation pressure in Fig. 9 for 70 MeV and Fig. 10 for 140 MeV irradiations. Error bars indicate the 95% confidence intervals of the fitted parameters.

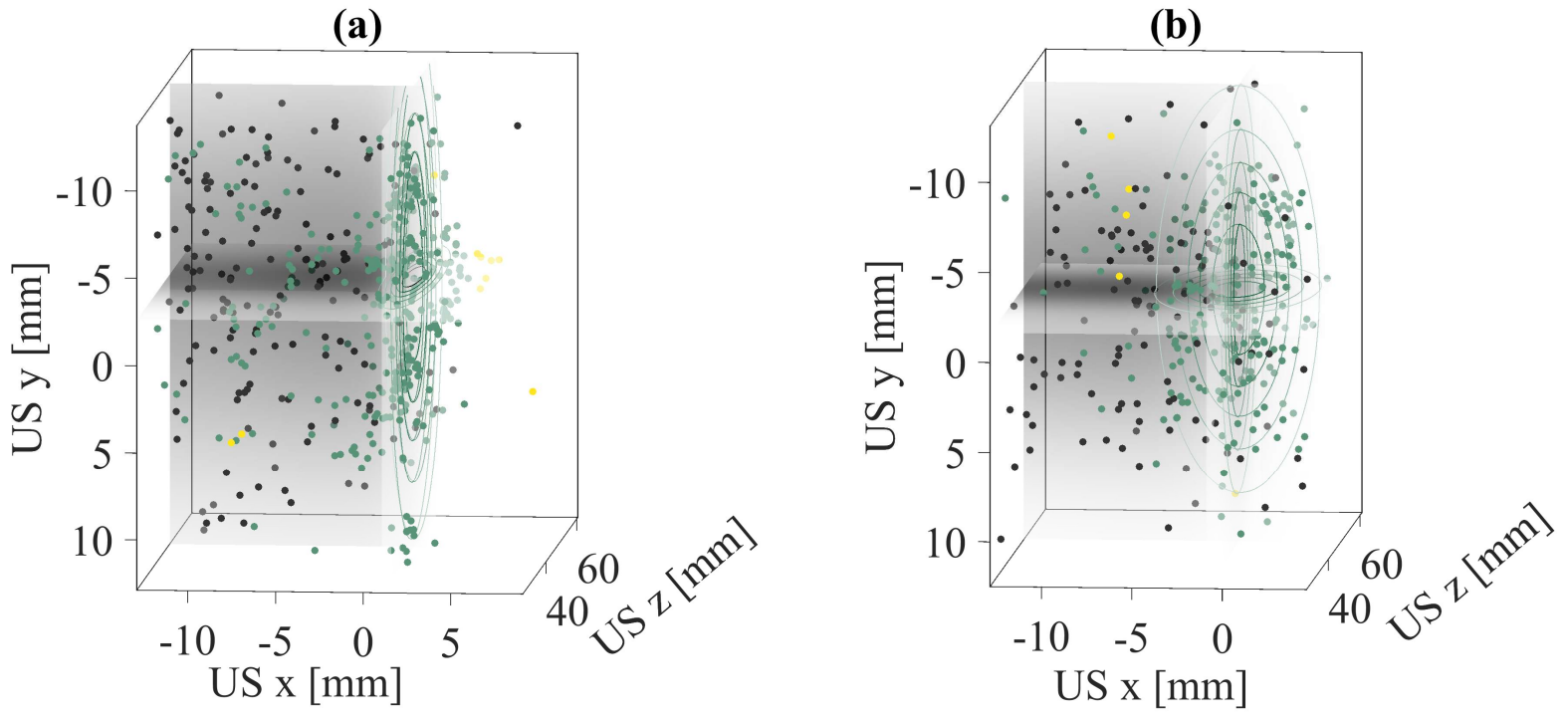

Fig 6: Spatial distribution of localized ND vaporization events (dots) overlaid on the distribution of protons stopping due to nuclear reactions (gray colormap) and due to Coulombic interactions at the end of their range (green contours), for 70 MeV (a) and 140 MeV (b) irradiations. Gray dots: irradiation without acoustic modulation, green dots: irradiation with acoustic modulation ($P_{\mathrm{rms,av}}$ = 265 kPa), and yellow dots: no irradiation (control) with acoustic modulation ($P_{\mathrm{rms,av}}$ = 265 kPa).

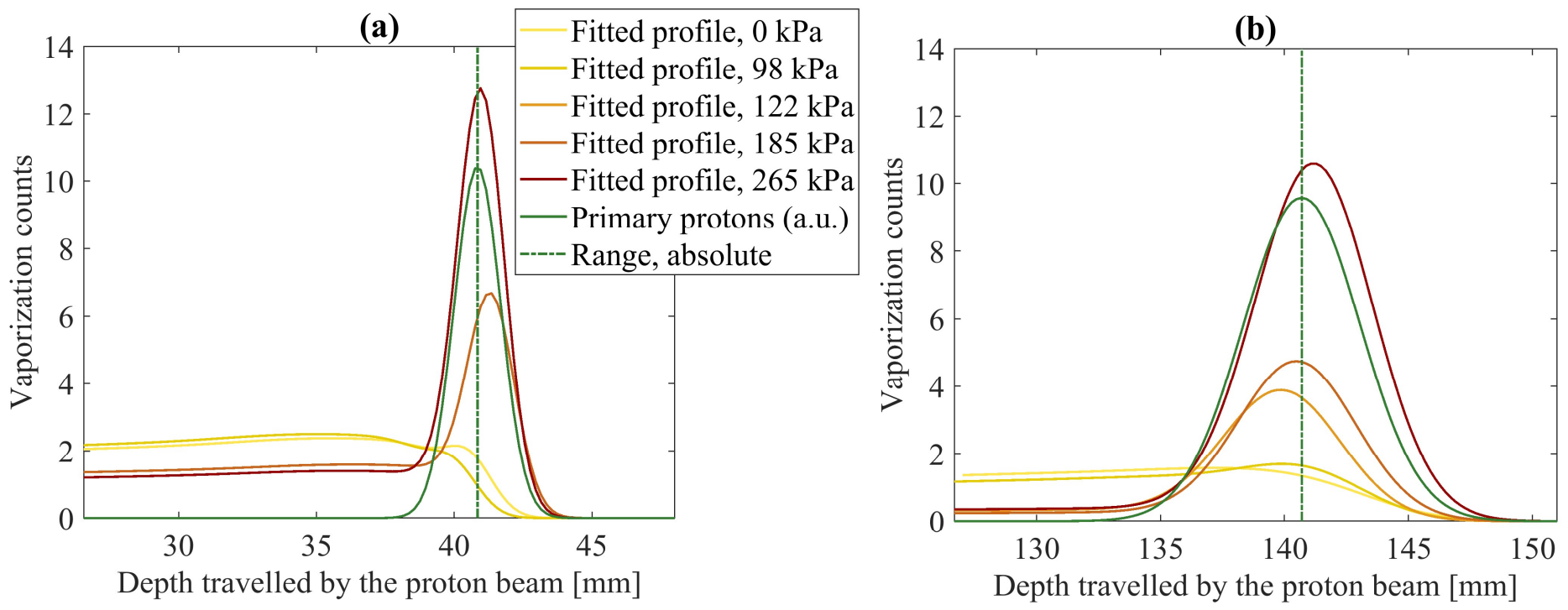

Fig 7: Fitted vaporization profiles in the proton beam direction (using equation 7) for phantoms irradiated using different acoustic modulation pressures at proton energies of 70 MeV (a) and 140 MeV (b). The expected distribution of primary protons stopping due to Coulombic interactions is displayed in green and peaks at the proton range (dotted green line). Reported pressure correspond to $P_{\mathrm{rms,av}}$ values.

The most clinically-relevant metric, the range shift, is shown in panels (a). At both energies, increasing the acoustic modulation pressure improved both the accuracy ($\mu_x$ approaching zero) and precision (narrower confidence intervals). For acoustic modulation pressures $P_{\mathrm{rms,av}} \geq 185$ kPa, all range estimates were within ±0.5 mm of the ground-truth value. A similar trend was observed for the estimated Bragg peak width $\sigma_x$, which was recovered within ±0.4 mm of the true value for pressures $P_{\mathrm{rms,av}} \geq$ 185 kPa.

In the directions lateral to the proton beam (US-y and US-z), the agreement appeared less accurate than in the range direction. In the US-y (elevation) direction, the estimated lateral shift and beam width were relatively stable across all acoustic modulation pressures, exhibiting no clear bias and remaining within ±2.5 mm of the ground truth. In contrast, the largest deviations were observed in the US-z (axial) direction, where the estimated shift differed from the reference by up to 7.8 mm and the error in the fitted beam width was up to 9 mm.

### F. Vaporization yield compared to theoretical predictions

For phantoms irradiated using acoustic modulation pressures $P_{\mathrm{rms,av}} \geq 265$ kPa, the normalized number of vaporizations events localized within the Bragg peak corresponded to 1.5% at 140 MeV and 2.5-3% at 70 MeV of the theoretical vaporization yield predicted from the proton fluence and ND concentration.

## IV. Discussion

In this contribution, we have shown that a spatially-uniform, wide acoustic modulation field can be generated and interleaved with vaporization monitoring on a single matrix array, offering a clinically-relevant, compact setup to achieve proton-induced DFB ND vaporization at body temperature. These results lift some major limitations of our previous proof-of-concept study [35], which used a single-element focused transducer for acoustic modulation and a linear array for vaporization detection: as acoustic modulation could only be applied to a narrow region, sensitization to protons was assessed by evaluating the number of vaporizations in a region distal to the Bragg peak, where vaporizations by secondary particles are expected to be negligible. In contrast, simultaneous volumetric acoustic modulation and imaging allows sensitization of NDs to protons across the whole field of view while monitoring treatment delivery in 3D, enabling proton range verification. While in the absence of acoustic modulation, vaporizations were induced by high-LET secondary particles (proximal, plateau profile), simultaneous acoustic modulation and proton irradiation enabled proton-induced vaporization (Gaussian peak at the proton range, Fig. 7). The required rms pressures (in the range 122-354 kPa) were well below the ADV thresholds in the MPa range typically reported in the literature [51], [52], [53], [54].

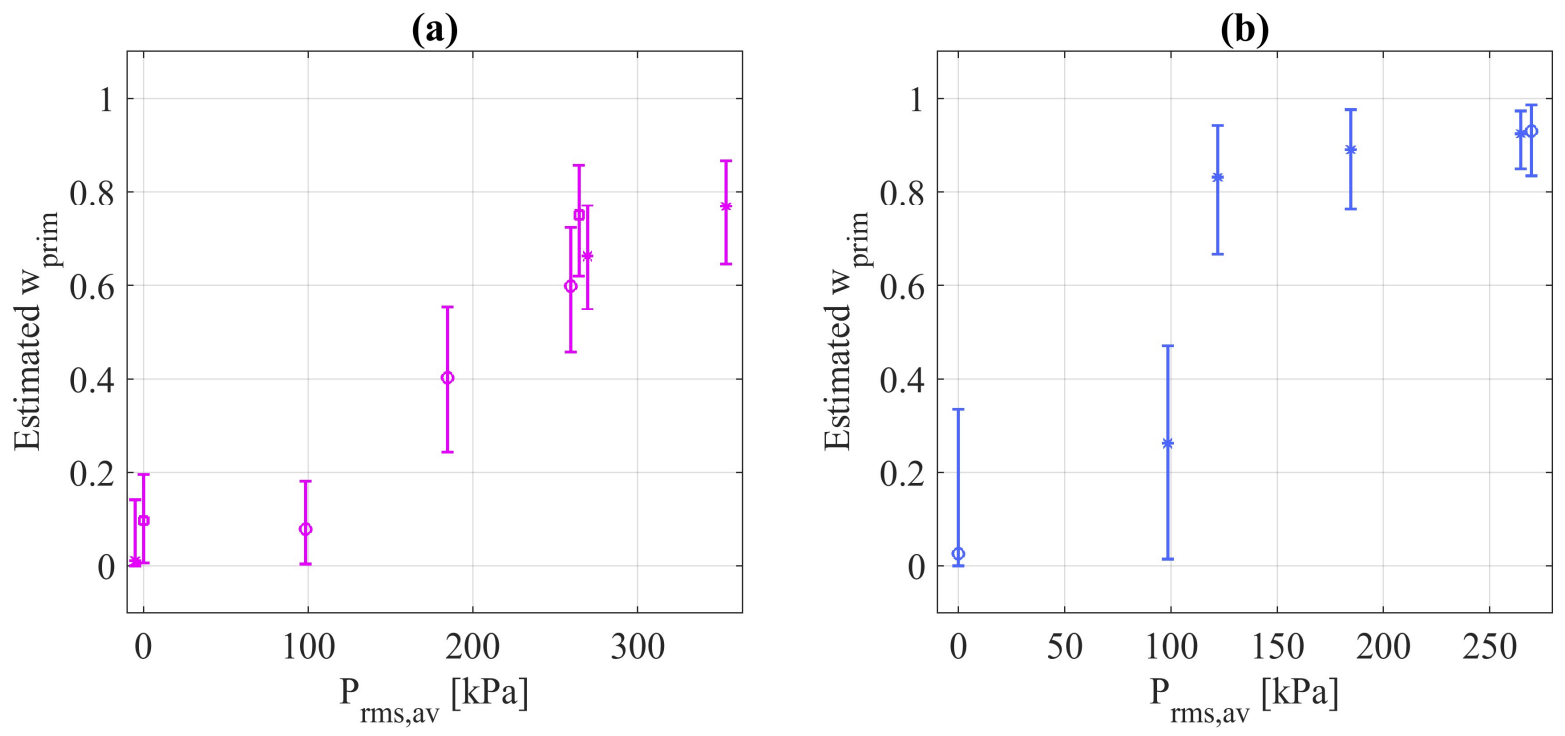

Fig 8: Fitted $w_{\mathrm{prim}}$ values as a function of the acoustic modulation pressure for phantoms irradiated with 70 MeV (a) or 140 MeV (b) protons. Different marker shapes indicate different phantoms, while same marker shapes correspond to successive irradiations of the same phantom. Vertical bars represent the 95% confidence intervals on the fitted values. A $w_{\mathrm{prim}}$ value of 1 means that all vaporizations were attributed to primary protons stopping due to Coulombic interactions at the end of their range. Of note, data points acquired at the same modulation pressures were positioned slightly apart to improve readability.

To ensure homogeneous sensitization throughout the irradiated volume, an optimization procedure was carried out using Field II simulations to identify a combination of four transmissions yielding the most homogeneous time-averaged pressure field. Despite inhomogeneous individual transmissions (Fig. 3a-d), both the PNP (whether NDs will be sensitized, Fig. 3e) and the time-averaged rms pressure (indicative of the ND vaporization probability, Fig. 4) were homogeneous across a period of 4 ms (4 transmissions at 1 kHz). This 4 ms averaging period is negligible compared to the “beam on” time of a typical proton therapy delivery, which is in the order of several seconds [55].

Due to the limited steering capabilities of the large-element matrix array used in this study, the homogeneous acoustic modulation region was restricted to approximately 18 × 18 $mm^2$, smaller than the array aperture (30 × 30 $mm^2$). Along depth, the use of quasi-plane waves (focus depth = 250 mm) ensured acoustic homogeneity in a 20 mm region centered on the proton beam axis (Fig. 4 and 5d). These dimensions were sufficient for the pristine proton beams used in this study, allowing 58.5% at 70 MeV and 78% at 140 MeV of the stopping protons to be able to interact with acoustically-modulated NDs (Fig. 5). To cover clinically-relevant treatment plans, wider acoustic modulation fields might be required and could be achieved using diverging waves with phased matrix arrays, or large element arrays with a wider aperture [56].

Acoustic modulation changed the vaporization response from one dominated by nuclear reaction products to one dominated by primary protons stopping at the end of their range (Fig. 7). This transition is reflected in the fitted values of $w_{\text{prim}}$, the fraction of vaporizations due to primaries, which displayed a sigmoid transition from ~0 (negligible vaporization by primaries) to around 0.75 at 70 MeV and 0.93 at 140 MeV as the acoustic modulation pressure increased (Fig. 8). These plateau values are comparable to those observed for irradiations at 50°C, a temperature at which the LET threshold is sufficiently low for ND vaporization by primary protons [57]. This emergence of a proton-induced vaporization peak strongly improved the accuracy and precision of range ($\mu_x$) and Bragg peak width ($\sigma_x$) estimation ( Fig. 9-10 a-b). Consistent with this observation, the range shift retrieval test was only successful (within 0.16 mm of the ground truth) when acoustic modulation was applied. Two factors likely contribute to this improvement. First, proton-induced vaporizations produce a localized Gaussian distribution that can be fitted with greater confidence than the broad plateau-like distribution induced by high-LET secondary particles. Second, acoustic modulation increased the number of detected vaporization events, thereby improving the statistical robustness of the fit. In addition, the shape and location of the secondaries distribution depend on nuclear reaction modelling, which is itself affected by uncertainties in the underlying cross sections [58]. By contrast, the proton-induced vaporization peak directly coincides with the proton range. In directions transverse to the proton beam, distribution fitting was less impacted by acoustic modulation as vaporizations originating from both primary and secondary particles were projected onto the same profile, effectively merging their contributions.

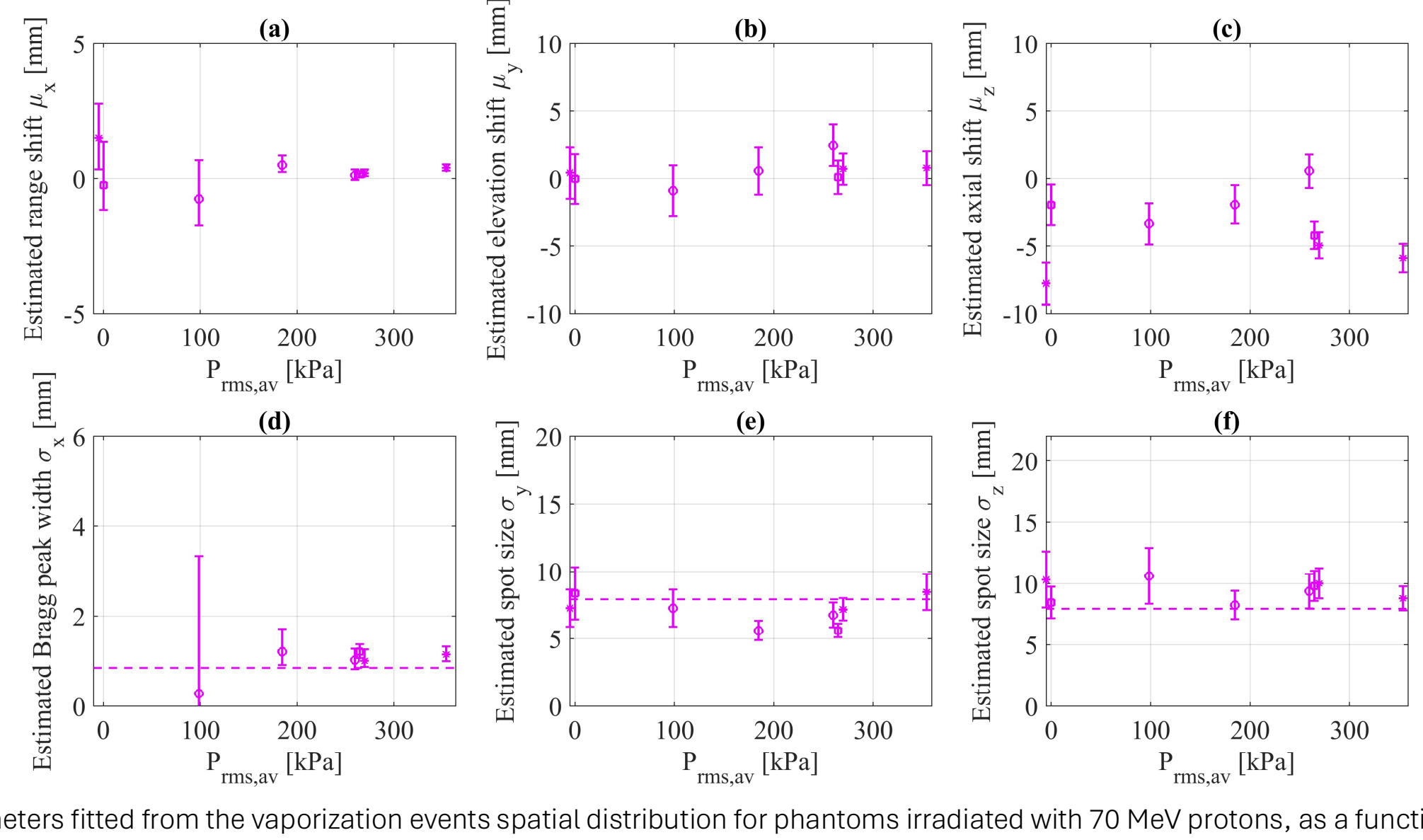

Fig 9: Beam parameters fitted from the vaporization events spatial distribution for phantoms irradiated with 70 MeV protons, as a function of the applied acoustic modulation pressure. (a-c) Fitted shifts of the beam centroid and (d-f) fitted Gaussian beam widths (σ) in the US-x (proton range), US-y (proton lateral), and US-z (proton lateral) directions. Dotted lines in (d-f) represent reference values from the beam model. Different marker shapes indicate different phantoms, while same marker shapes correspond to successive irradiations of the same phantom. Vertical bars represent the 95% confidence intervals on the fitted values. Of note, data points acquired at the same modulation pressures were positioned slightly apart to improve readability.

A vaporization response to primary protons was observed for spatially-averaged rms acoustic pressures $P_{rms,av}$ as low as 122 kPa (Fig. 7), corresponding to a PNP of 192 kPa. This threshold is lower than the approximately 250 kPa rms pressure previously reported using a focused single-element transducer [35]. This discrepancy could be explained by several factors: the reported pressures in [35] might have been overestimated as they correspond to the spatial peak pressure at the transducer focus; while the sensitization success criterion was based on reaching an (arbitrary) vaporization count over a larger area. Alignment inaccuracies between the linear imaging array and the focused single-element transducer might have impacted the actual pressure in the imaged plane. Finally, the sensitization threshold might depend on the ultrasound frequency, which differed in the two studies (1.1 MHz in [35] and 2 MHz here).

In both studies, the measured sensitization pressures exceeded the quasi-static pressure decrease of 50 – 90 kPa predicted by the thermal spike theory for proton sensitization at body temperature. This theoretical range corresponds to the condition where the LET threshold equals the maximum LET reached by a proton near the end of its range (74.5 keV/µm in DFB [59]). However, under this condition the proton reactive length, defined as the portion of the proton track for which the LET exceeds the vaporization threshold [45], is effectively zero. Consequently, the probability of proton-induced vaporization is also expected to approach zero. Further reductions in the LET threshold require larger pressure reductions, which increase the reactive length and therefore the vaporization probability. As a result, experimentally determined sensitization pressures are expected to exceed the theoretical 50-90 kPa range. Additionally, the thermal spike theory assumes quasi-static conditions, while dynamic pressure variations in the MHz range might lead to competition between the growth rate of vapor embryos formed by protons during the rarefactional phase and their compression during the consecutive compressional phase. This effect might lead to vapor embryos recondensation and hence prevent complete ND vaporization. Therefore, increased modulation frequencies could yield higher thresholds for proton sensitization compared to quasi-static conditions. Finally, the thermal spike theory does not account for the shell viscoelastic properties or potential Laplace pressure effects. Modelling these effects and optimizing the center frequency to maximize the vaporization yield was outside of the scope of this study. Our choice of 2 MHz was a compromise between the need to achieve sufficient image resolution while ensuring NDs remained in the rarefactional phase for a sufficiently long time.

Consistent with this interpretation, the vaporization yields observed in this work remained substantially lower (by nearly two orders of magnitude) than those previously reported using passive imaging at 50°C [57] (*i.e.* when NDs are continuously sensitive to primary protons), which agreed with the analytic prediction of 15.7 events per $10^8$ protons and µM NDs [45]. As the vaporization probability is proportional to the fraction of time during which NDs experience pressures below the sensitization threshold, lower yields are inevitable compared to conditions in which NDs are continuously sensitive to protons. In addition, limitations of the current vaporization detection pipeline likely resulted in undetected vaporization events, as weak vaporizations may have been suppressed by high-pass filtering or excluded by the ULM detection threshold.

This study has several limitations. The ND concentration was intentionally kept low to maintain sufficiently sparse vaporization events for super-localization. In counterpart, this low number of events (~ 100-400 per irradiation) limited the accuracy of proton distribution fitting in the beam transverse directions (Figs 9-10 b-c and e-f). Consequently, the confidence intervals on the fitted parameters were high (several mm). While vaporization monitoring was shown to be feasible across the entire field of view, we could therefore neither entirely validate that the vaporization distribution accurately follows the lateral proton dose falloff, nor exclude potential heterogeneities in the vaporization probability affecting vaporization distribution in the US-z and US-y directions. In the proton beam direction (US-x), the effect of local pressure inhomogeneities seemed to be negligible as the vaporization distribution closely matched the proton stopping distribution (Fig. 7) and both range and beam width estimates were very close to ground truth values (Figs 9-10 a-b). Future studies will aim at reaching higher vaporization densities and assessing the global spatial conformity between the proton distribution and vaporization maps using tools such as the gamma index.

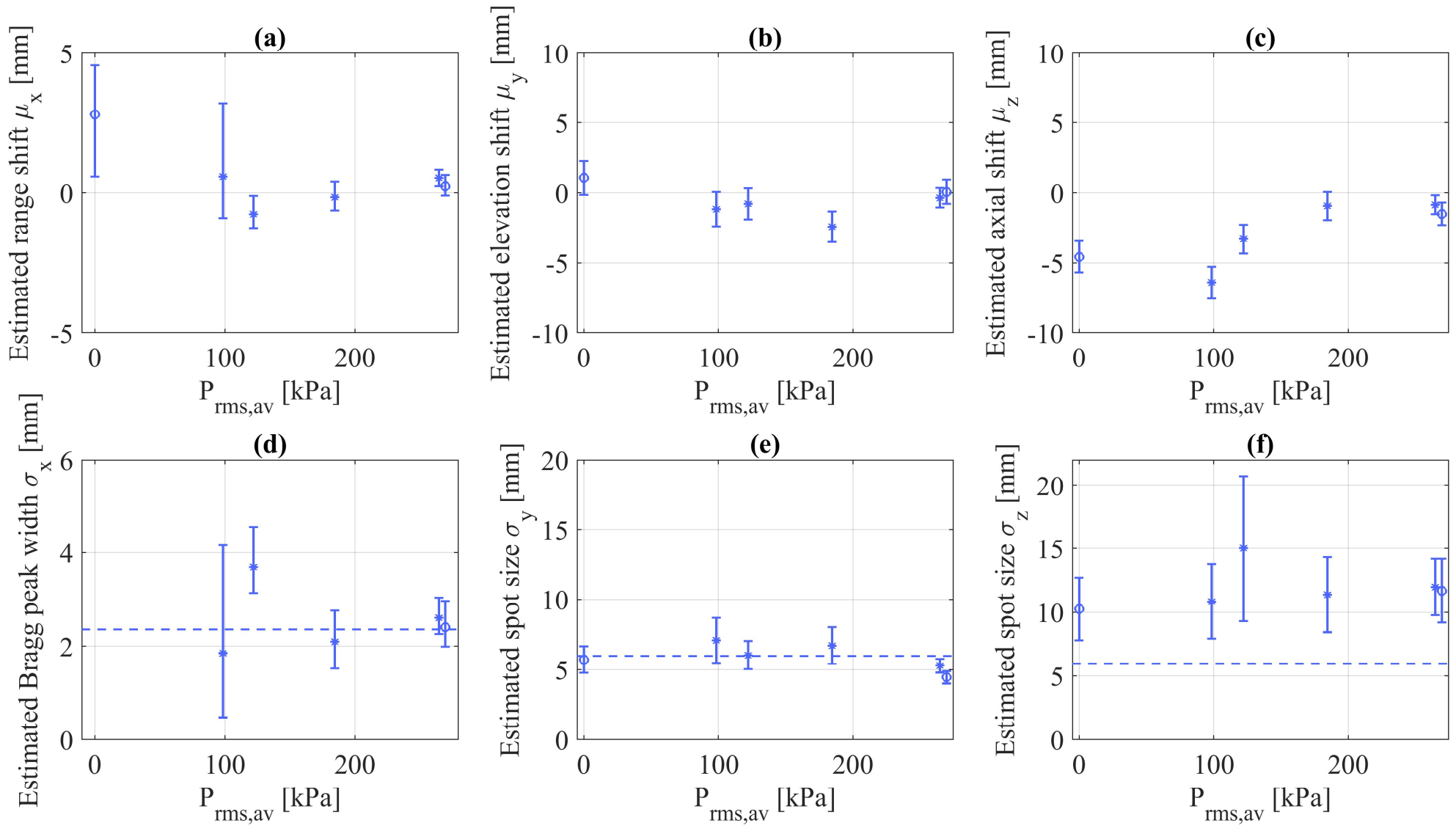

Fig 10: Beam parameters fitted from the vaporization events spatial distribution for phantoms irradiated with 140 MeV protons, as a function of the applied acoustic modulation pressure. (a-c) Fitted shifts of the beam centroid and (d-f) fitted Gaussian beam widths (σ) in the US-x (proton range), US-y (proton lateral), and US-z (proton lateral) directions. Dotted lines in (d-f) represent reference values from the beam model. Different marker shapes indicate different phantoms, while same marker shapes correspond to successive irradiations of the same phantom. Vertical bars represent the 95% confidence intervals on the fitted values. Of note, data points acquired at the same modulation pressures were positioned slightly apart to improve readability.

Multiple irradiations (up to four, see Table I) were performed on the same phantom because of limited beamtime availability. Therefore, a decrease of the ND concentration over time due to both ND gradual dissolution and radiation-induced vaporization to MBs could have occurred, preventing accurate quantitative comparisons between vaporization counts for different irradiations.

Vaporization detection relied on an in-house 3D ULM pipeline and assumed that the beginning of each microbubble track corresponded to a vaporization event. While the obtained vaporization maps agree well with the proton distribution, we observed labelled vaporization events outside of the irradiation duration, suggesting limitations in the pipeline's specific detection of ND vaporizations. The MB detection threshold was adapted to limit the percentage of those false positives to at most 5%. Nevertheless, the reported number of vaporizations might deviate from the true radiation-induced vaporization count due to (i) undetected vaporizations whose intensity was below the fixed MB detection threshold, (ii) residual noise inaccurately labelled as vaporizations, (iii) spontaneous vaporizations, (iv) tracking errors *i.e.* a previous MB track being labelled as a new vaporization.

To increase the accuracy and reliability of the vaporization maps, several options should be explored. This study was implemented on a large-element matrix array to ensure that sufficiently high acoustic pressures could be reached across a wide field of view. However the large element size with respect to the wavelength (3.6λ) led to a degraded image quality (in particular, strong grating lobes) compared to fully-populated arrays. As we showed that proton sensitization takes place at moderate rms pressures (122 kPa), phased matrix arrays with a more conventional λ/2 pitch might achieve sufficient pressures with plane or diverging waves, leading to improvements both in the acoustic modulation homogeneity and in image quality. On the other hand, large-element arrays offer a wide field of view with a reasonable channel count (and hence less expensive hardware), and enable whole-organ volumetric imaging with high sensitivity [56]. The addition of diverging lenses can drastically increase the element directivity [60], [61], and thereby improve both the acoustic modulation homogeneity and the image quality. The latter could be further enhanced by increasing the number of compounded angles, or through advanced beamforming strategies.

In parallel, ND vaporization detection should be optimized. Our current active imaging approach of detecting resulting MBs relies on 3D ULM algorithm performance, in particular the high-pass filtering step. The latter is necessary to keep MBs sparse as irradiation progresses, but complicates MB tracking by suppressing slowly-moving MBs and introducing oscillatory responses. In addition, freely-dispersed MBs were subject to different flow conditions depending on the intensity of the acoustic radiation force, proportional to the acoustic modulation pressure. This resulted in different filter settings being required depending on the observed flow, which further limits quantitative comparisons between phantoms. Deep-learning approaches could be leveraged to address these limitations and improve the specificity of ND vaporization detection [62]. An alternative option would be to detect the acoustic signatures generated by vaporization events (*i.e.* passive ULM [57]); however, isolating those weak signals from the high acoustic scatter induced by the modulation wave is challenging.

The freely-flowing NDs setup used in this study is not fully representative of *in vivo* conditions, where NDs may experience variable flow and vascular confinement conditions. The presence of tissue will require clutter filtering, possibly combined to the use of nonlinear contrast imaging sequences. Furthermore, spatially varying tissue attenuation could reduce the homogeneity of the acoustic modulation field, while vessel confinement might affect the ND vaporization threshold [63]. Conversely, the more organized intravascular ND flow may benefit the ULM processing pipeline by facilitating the separation of coherent MB signals from background noise and improving MB tracking performance compared with free-flow conditions. Therefore, *in vivo* studies remain an important next step to confirm the clinical potential of acoustic modulation for proton range verification.

To assess the suitability of such acoustic modulation sequences for *in vivo* translation, acoustic safety parameters were derived from hydrophone measurements and compared with regulatory limits [64], [65], [66]. For an acoustic modulation pressure $P_{\mathrm{rms,av}}$ of 265 kPa, the mechanical index (MI) was 0.4, well below the limit of 1.9. The thermal index in soft tissue (TIS) had a high value of 10, exceeding the recommended limit of 6. This elevated TIS primarily reflects the large insonified field and associated total acoustic power, rather than high local acoustic intensities. Accordingly, the spatial-peak pulse-average intensity ($I_{\mathrm{sppa}}$) and spatial-peak temporal-average intensity ($I_{\mathrm{spta}}$) were 9 W/cm$^2$ and 221 mW/cm$^2$, respectively, both well below their respective regulatory limits of 190 W/cm$^2$ and 720 mW/cm$^2$. Assuming a soft tissue absorption coefficient of 0.5 dB/cm/MHz, the estimated tissue heating rate is 0.01°C/s. This suggests that the proposed modulation sequence might be compatible with treatment durations in proton therapy (~ 1 minute).

## V. Conclusion

In this paper, we interleaved volumetric acoustic modulation and ND vaporization monitoring on a large element matrix array synchronized with the delivery of a pulsed proton beam. This enabled ND vaporization by primary protons at body temperature across the irradiated volume and treatment verification using the 3D spatial distribution of ND vaporization events. Future studies will focus on *in vivo* translation and implementation on phased matrix arrays to improve both the modulation field size and homogeneity and the image quality.

## VI. Acknowledgment

The authors thank the IBA Leuven team for their help in synchronizing acoustic acquisitions with pulsed beam delivery, and Mr Quentin Flandroy for his assistance with the interpretation of the proton therapy machine delivery log files.